\documentclass[12pt]{article}

\usepackage[a4paper,
            twoside = true,
			top = 30mm,
		    bottom = 30mm,
            inner = 25mm,
            outer = 25mm
		    ]{geometry}

\usepackage{amsmath,amsfonts,mathtools,amssymb,slashed,euscript,bm,bbm,graphicx}
\usepackage[table,dvipsnames]{xcolor}
\usepackage[labelfont=bf,width=0.95\textwidth]{caption}
\usepackage{cite}
\usepackage{cancel}
\usepackage[
    pdfencoding=unicode,
    bookmarksnumbered=true,
    urlcolor=blue,
    linkbordercolor=red,
    citebordercolor=green,
    bookmarksopen=true
    ]{hyperref}
\usepackage{ytableau}
\usepackage{simpler-wick}
\usepackage{booktabs}
\usepackage[table]{xcolor}
\usepackage{array}

\definecolor{rulegray}{gray}{0.72}
\definecolor{headergray}{gray}{0.92}

\renewcommand{\arraystretch}{1.25}  
\numberwithin{equation}{section}    
\allowdisplaybreaks                 
\newcommand{\pair}[2]{#1\,{\footnotesize (#2)}}
\newcommand{\nm}[1][]{n_{#1-}}
\newcommand{\nms}[1][]{\spac/\hspace{-2.3mm}n_{#1-}}
\newcommand{\np}[1][]{n_{#1+}}
\newcommand{\nps}[1][]{\spac/\hspace{-2.3mm}n_{#1+}}
\newcommand{\spac}{{\hspace{0.3mm}}}

\newcommand{\lag}{{\mathcal{L}}}
\newcommand{\loc}{\mathrm{loc}}
\newcommand{\rad}{\mathrm{rad}}

\title{On the universality of soft emission beyond\texorpdfstring{\\[2mm]}{ }the next-to-soft order
}

\begin{document}

\begin{titlepage}

\begin{flushright}
{\small
TUM-HEP-1614/26\\
September~9, 2026\\
}
\end{flushright}

\makeatletter
\vskip0.8cm
\pdfbookmark[0]{\@title}{title}
\begin{center}
{\Large\bf\boldmath \@title}
\end{center}
\makeatother

\vspace{0.5cm}
\begin{center}
\textsc{Martin~Beneke}, 
\textsc{Maria~Santana},
and \textsc{Michel~Stillger}\\[6mm]
\textit{Physik Department T31, Technische Universit\"at M\"unchen\\
James-Franck-Stra\ss e 1, D-85748 Garching, Germany}
\end{center}

\vspace{0.6cm}
\pdfbookmark[1]{Abstract}{abstract}
\begin{abstract}
\noindent 
The Low-Burnett-Kroll theorem (extended to non-abelian gauge theory) states that the tree-level emission amplitude of a soft gauge boson from an arbitrary hard process can be expressed in terms of the non-radiative amplitude including the next-to-soft term. Reformulating the soft theorem in the framework of an on-shell effective theory leads to a natural gauge-invariant split into an external and internal emission process to any order in the soft expansion. In this framework, we derive a compact expression for the all-order external emission amplitude expressed in terms of the non-radiative amplitude, and express the local emission amplitude through a single gauge-invariant operator at every order beyond the next-to-soft one. For scalar QCD, we compare the number of Lorentz- and $SU(3)$-invariant amplitudes for the radiative and non-radiative hard process with an arbitrary number of scalar quarks and antiquarks.
\end{abstract}

\thispdfpagelabel{t}
\end{titlepage}

\pagenumbering{arabic}
\setcounter{page}{1}

\section{Introduction}
\label{sec:S1_introduction}

In the limit where the momentum of an emitted massless particle becomes small, scattering amplitudes exhibit universal factorization properties known as soft theorems.
These theorems express the amplitude in terms of soft operators acting on the corresponding non-radiative amplitude.
In QED, the structure of soft radiation is well understood including the first subleading order in the soft expansion.
The next-to-soft theorem was first derived at tree level by Low~\cite{Low:1958sn} and Burnett and Kroll~\cite{Burnett:1967km}, and has recently been extended to all orders in the electromagnetic coupling, allowing for an arbitrary number of soft emissions~\cite{Engel:2023ifn,Engel:2023rxp}. At tree level, the soft theorem structure generalizes straightforwardly to non-abelian gauge theories, 
but beyond this order the structure becomes substantially more intricate due to gluon self interactions and non-trivial colour correlations between external legs~\cite{Czakon:2023tld}. 

The tree-level radiative amplitude $\mathcal{A}_{N+1}^\rad$ can be expressed in terms of emission from an external leg of an $N$-particle amplitude $\mathcal{A}_N$ and internal emission from the hard process in the form 
\begin{align} \label{eq:standardsplit}
    \mathcal{A}_{N+1}^\rad = \mathcal{A}_{N+1}^\mathrm{ext} + \mathcal{A}_{N+1}^\mathrm{int} \,.
\end{align}
Internal emission contributes from the next-to-soft order,
but the Ward identity relates it to the non-radiative amplitude, resulting in the well-known expression
\begin{align} \label{eq:LBK}
    \mathcal{A}_{N+1}^{\rad}(\{\underline{p}\},k) = - g_s \sum_{i=1}^N \bm{T}_i^a \bigg[ \frac{p_i\cdot\varepsilon^\ast(k)}{p_i\cdot k} + \frac{k_\mu \varepsilon^\ast_\nu(k) \spac J_i^{\nu\mu}}{p_i\cdot k} + \mathcal{O}(k) \bigg] \mathcal{A}_N(\{\underline{p}\}) \,,
\end{align}
where $k$ denotes the soft gluon momentum, $\bm{T}_i^a$ the colour operator in the representation of energetic particle $i$~\cite{Catani:1996jh,Catani:1996vz}, and $J_i^{\mu\nu}$ the angular-momentum operator.
Beyond the next-to-soft order in the expansion of the radiative amplitude, Ward identities do not constrain the radiative amplitude completely, and non-universal structures emerge. This raises the question of whether the degree of non-universality can be quantified order by order in the soft expansion, and whether a systematic classification of the operator structures governing both universal and process-dependent contributions can be established.

\begin{figure}[t]
    \centering
    \begin{tabular}{ccc}
        \includegraphics[scale=1]{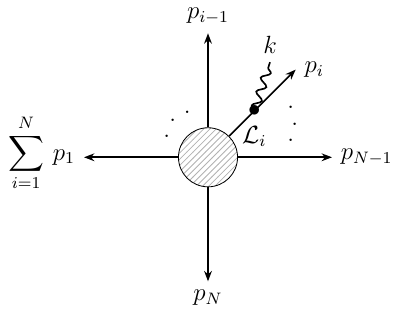} &\hspace*{12mm}& \includegraphics[scale=1]{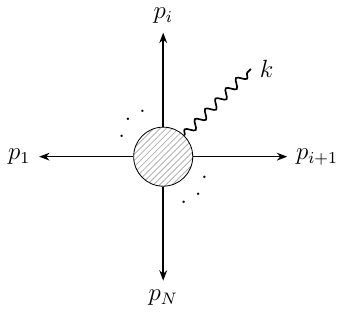}
    \end{tabular}
    \caption{Universal $\mathcal{A}_{N+1}^{\lag}$ (left) and non-universal part $\mathcal{A}_{N+1}^{\loc}$ (right) of the radiative amplitude. The insertion of the SCET Lagrangian $\mathcal{L}_i$ is denoted by a black dot.}
    \label{fig:radiative_amplitude_split}
\end{figure}

Working directly with full QCD does not provide an ideal setting for addressing these questions, as the external and internal amplitudes are not gauge-invariant. Instead, formulating the soft expansion in the framework of an on-shell soft-collinear effective theory (SCET)~\cite{Bauer:2000yr,Bauer:2001yt,Beneke:2002ph,Beneke:2002ni}, the radiative amplitude separates into (see Figure~\ref{fig:radiative_amplitude_split})
\begin{align} \label{eq:amplitude_split}
    \mathcal{A}_{N+1}^\rad = \mathcal{A}_{N+1}^\lag + \mathcal{A}_{N+1}^\loc\,,
\end{align}
where the first term arises from Lagrangian interactions and the second from local operators. Both terms are now separately gauge-invariant due to the emergent soft gauge symmetry of SCET and constrained by these symmetries. This perspective has already been used to rederive the soft theorems from SCET in both gauge theories and gravity~\cite{Beneke:2021umj}, showing that universality up to next-to-soft order in gauge theories and next-to-next-to-soft order in gravity follows simply from the absence of gauge-invariant operators that generate emissions from the hard vertex. In this article, we apply this approach to characterize the degree of non-universality by counting the independent operators that enter $\mathcal{A}_{N+1}^{\loc}$ at each order in the soft expansion. 

\section{Structure of the amplitude in SCET}
\label{sec:S2_rad_amplitude_SCET}

\paragraph{Non-radiative amplitude} Before turning to the radiative amplitude, we discuss the representation of the non-radiative one. The non-radiative process $0\to p_1+\dots+p_N$ generates $N$ energetic, on-shell, massless, \emph{scalar} quarks from a gauge-invariant source (shaded circle in the left Figure~\ref{fig:radiative_amplitude_split}). The momenta are decomposed as
\begin{align}
    p_i^\mu =\np[i] p_i \, \frac{\nm[i]^\mu}{2} + \nm[i] p_i \, \frac{\np[i]^\mu}{2} + p_{i\perp}^\mu \equiv (\np[i] p_i,\nm[i] p_i,p_{i\perp} ) \,,
\end{align}
where $n_{i\pm}^\mu$ are pairs of light-like vectors satisfying $\np[i]\nm[i]=2$.
We assume that the non-radiative amplitude represents a hard process, so that the directions of all $p_i$ are non-collinear, i.e.~$\nm[i]\nm[j]=\mathcal{O}(1)$ for $i\neq j$. The $p_i$ are roughly aligned with the reference vectors $\nm[i]$, with $p_i\sim(1,\lambda^2,\lambda)$ and $\lambda\ll 1$. With this counting, the soft momentum scales as $k\sim \lambda^2$. This set-up is known as SCET$_{\rm I}$, and it forms the starting point for the soft expansion in $\lambda^2$. 

Expanding in the small transverse momenta with respect to the reference axes, we express the \emph{on-shell} non-radiative amplitude as 
\begin{align} \label{eq:nr}
    \mathcal{A}_{N}(\{\underline{p}\}) &= \langle \{\underline{p}\} | \sum_M C^{AM}\ast J^{AM} |0\rangle
    \nonumber\\[-2mm]
    &= \sum_{m_1=0}^\infty \ldots \sum_{m_N=0}^\infty (p_{1\perp})^{m_1} \ldots (p_{N\perp})^{m_N} \, C^{Am_1\dots Am_N}(\{\underline{\np p}\}) \,,
\end{align}
where $M=(m_1\dots m_N)$ is a multi-index.
We use the on-shell conditions to eliminate $\nm[i] p_i = -p_{i\perp}^2/\np[i] p_i$ and regard the amplitude as a function of the transverse and $\np[i] p_i$ components only. In SCET, this expansion has an operator representation in terms of so-called $(Am)$-type collinear currents,
\begin{align} \label{eq:collinearcurrents}
    &J^{AM}(\{\underline{t}\}) = \prod_{i=1}^N J_i^{Am_i}(t_i) = \prod_{i=1}^N (-i)^{m_i} (\partial_{i\perp}^{\mu_1}\dots\partial_{i\perp}^{\mu_{m_i}} \, \bar{\chi}_i)(t_i\np[i]) \,,
\end{align}
where $\chi_i$ is the collinear quark field that describes (anti-)quark fields moving along $\nm[i]$. The $C^{AM}$ encode the short-distance physics and are determined by on-shell matching to the corresponding full-theory non-radiative amplitude on the left-hand side of~\eqref{eq:nr}. In coordinate-space operator language 
\begin{align} \label{eq:conv}
    C^{AM} \ast J^{AM}\equiv \int\!d^Nt \, C^{AM}(\{\underline{t}\}) \,J^{AM}(\{\underline{t}\}) 
\end{align}
denotes convolution over the light-cone positions $\{\underline{t}\}=\{t_1,\dots,t_N\}$. To improve readability, Lorentz, Dirac, and colour indices are left implicit here. 

Since the (small) amount of misalignment of the reference vectors $n_{i\pm}$ with respect to their $p_i$ is arbitrary, Lorentz invariance, rephrased as reparametrization invariance in SCET~\cite{Manohar:2002fd}, relates \emph{all} short-distance coefficients $C^{Am_1\dots Am_N}$ to the single coefficient $C^{A0}\equiv C^{A0\dots A0}$.
The explicit form of this relation, which appears not to have been given before, is presented in Appendix~\ref{app:CAM}.

\paragraph{Universal Lagrangian insertion amplitude} Turning to the radiative amplitude, the emission of a soft gluon from an external leg of the $N$-particle amplitude is given by the matrix element of the time-ordered product
\begin{align} \label{eq:tproduct_uni}
    &\mathcal{A}_{N+1}^{\lag}(\{\underline{p}\},k) \equiv \langle \{\underline{p}\},k | \, i\!\int\!d^4x \; \text{T} \, \bigg\{\sum_M C^{AM} \ast J^{AM},\, \sum_{i=1}^N \mathcal{L}_i(x)\bigg\}|0\rangle \,.
\end{align}
Several important comments can be made. (1) Since collinear modes of different directions $\nm[i]$ interact only via exchange of soft modes, the SCET Lagrangian is of the form
\begin{align}
    \mathcal{L}_\mathrm{SCET} = \sum_{i=1}^N \mathcal{L}_i(\psi_{i},\psi_s) +\mathcal{L}_s(\psi_s) \,,
\end{align}
with $N$ copies of the soft-collinear interaction Lagrangian. Accordingly, there is an enlarged collinear and soft gauge symmetry $\prod_{i=1}^N SU(3)_i\otimes SU(3)_{s}$, where the collinear fields transform in a soft background~\cite{Beneke:2002ni}. The Lagrangian can be written down exactly to all orders in the $\lambda$ expansion~\cite{Beneke:2002ni} and the Lagrangian interactions are tree-level exact to all orders in $\lambda$~\cite{Beneke:2002ph}. (2) The short-distance coefficients $C^{AM}$ are the same \emph{on-shell} matching coefficients that appear in the non-radiative amplitude. The Lagrangian insertion amplitude is manifestly gauge-invariant by itself and obviously universal, i.e. it depends only on the non-radiative hard process and no other. (3) $\mathcal{A}_{N+1}^{\lag}$ is not the same as $\mathcal{A}_{N+1}^{\rm ext}$ that appeared in~\eqref{eq:standardsplit}, since the SCET Lagrangian is designed to reproduce QCD amplitudes only on-shell. Any effect of the small off-shellness of the momentum $p_i+k$, which connects the Lagrangian vertex to the non-radiative process, goes into the local amplitude $\mathcal{A}_{N+1}^\loc$ by on-shell matching. This is precisely what makes both parts, $\mathcal{A}_{N+1}^\lag$ and $\mathcal{A}_{N+1}^\loc$, separately gauge-invariant. 

The computation of the radiative amplitude can be considerably simplified by  choosing the reference vectors $\nm[i]$ to be aligned with the external collinear momenta $p_i$, such that
\begin{align} \label{eq:frame}
    p_i^{\mu} = \np[i] p_i \, \frac{\nm[i]^\mu}{2} \,,
    \quad
    i=1,\ldots,N 
\end{align}
exactly. With this choice, every operator $J_i^{Am_i}$ appearing in the time-ordered product defining the universal part of the radiative amplitude~\eqref{eq:tproduct_uni} can be reduced to the $(A0)$ operator whenever it appears together with a Lagrangian insertion $\mathcal{L}_j$ from a different collinear sector, $i\neq j$.
Indeed, at tree level 
\begin{align}
    \langle p_i | J_i^{Am_i}(t_i) | 0 \rangle = p_{i\perp}^{\mu_1} 
    \ldots p_{i\perp}^{\mu_{m_i}} \, e^{it_i \np[i] p_i} \,,
\end{align}
which vanishes for $m_i\neq 0$ by our choice~\eqref{eq:frame}.
This allows us to express the universal part of the tree-level radiative amplitude as 
\begin{align} \label{eq:ALNp1}
    \mathcal{A}_{N+1}^{\lag}(\{\underline{p}\},k) = \sum_{i=1}^N \sum_{m_i=0}^\infty C_i^{Am_i} \ast \langle p_i,k | \, i\!\int\!d^4x \, \text{T}\big\{ J_i^{Am_i}, \mathcal{L}_i(x) \big\} |0\rangle \prod_{j\neq i}^N \,\langle p_j | J_j^{A0} | 0\rangle \,,
\end{align}
where we introduced the shorthand notation $C_i^{Am_i}\equiv C^{A0\dots Am_i\dots A0}$.
The problem is thus reduced to calculating the one-jet radiative amplitude
\begin{align} \label{eq:Ai1}
    \mathcal{A}_{i+1}^{\lag}(p_i,k) = \sum_{m_i=0}^\infty C_i^{Am_i} \ast \langle p_i,k | \, i\!\int\!d^4x \, \text{T} \big\{ J_i^{Am_i}, \mathcal{L}_i(x) \big\} |0\rangle \,.
\end{align}
We will compute~\eqref{eq:tproduct_uni} in Section~\ref{sec:S3_uni}. 

\paragraph{Local emission amplitude} Direct emission of a soft gluon from the hard process (right Figure~\ref{fig:radiative_amplitude_split}) is represented by matrix elements of local-emission operators
\begin{align} \label{eq:local_emission}
   & \mathcal{A}_{N+1}^{\loc}(\{\underline{p}\},k)
    \equiv \langle \{\underline{p}\},k |\sum_Y C_\loc^Y \ast J_\loc^Y|0\rangle \,,
\end{align}
where $J_\loc^Y$ contains a soft field and $C_\loc^Y$ denotes the corresponding process-dependent short-distance coefficient, which defines a new hard amplitude different from the original non-radiative one. Note that ``local'' means that the soft gluon is emitted from a hard process---for the choice~\eqref{eq:frame} of reference vectors, the operators themselves take the form
\begin{align} \label{eq:Jlocal}
    J_\loc^Y(\{\underline{t}\}) = J_s^Y(0) \prod_{i=1}^N \bar{\chi}_i(t_i \np[i]) \,, 
\end{align}
and are non-local along the light-cones of the $N$ energetic, massless particles. 
Here $J_s^Y$ is a gauge-covariant soft operator (such that $J_\loc^Y$ is invariant) localized at the origin. 
The local-emission operators relevant to tree-level emission of a single gluon are 
\begin{align} \label{eq:js} 
    J_s^{(n+2),\,\rho_1\dots\rho_{n} \,\mu\nu }(0) = [-iD_s^{\rho_1},[-iD_s^{\rho_2},\ldots,[-iD_s^{\rho_{n}},ig_sF_s^{\mu\nu}]\ldots]](0) \sim k^{n+1} 
\end{align}
for $n=0,1,\ldots$, where $iD_s^\rho=i\partial^\rho+g_s\spac A_s^\rho$ and $ig_s F_s^{\mu\nu} =[iD_s^\mu,iD_s^\nu]$ with $g_s$ being the strong coupling constant. 
The notation is such that $J_s^{(k)}$ contributes at the (next-to)${}^{k}$-soft  order. 
We highlight that at every order in the soft expansion starting from the next-to-next-to-soft order $\mathcal{O}(k)$, there is only a single operator structure as a consequence of manifest gauge invariance of $\mathcal{A}_{N+1}^{\loc}$.
More precisely, since we work with collinear-gauge-invariant building blocks $\chi_i$, collinear gauge invariance is automatically satisfied for the operators~\eqref{eq:Jlocal}.
Soft gauge invariance, however, requires the $\chi_i$ and $J_s^Y$ to combine into colour singlets.
Together with Lorentz invariance, this constraint will be used to determine the number of independent non-universal amplitudes $C_\loc^Y$ in Section~\ref{sec:S4_non_uni}.


\section{Universal Lagrangian emissions}
\label{sec:S3_uni}

As discussed in the previous section, it is enough to determine the radiative amplitude $\mathcal{A}_{i+1}^{\lag}$ for a single collinear sector.
We split the SCET Lagrangian $\mathcal{L}_i = \mathcal{L}_i^{(2)} + \mathcal{L}_i^\mathrm{int}$ into a part containing the kinetic terms and one containing interactions.
Omitting the index $i$, the interaction Lagrangian relevant to single soft gluon emission reads~\cite{Beneke:2021aip}
\begin{align} \label{eq:lag}
    \mathcal{L}^\mathrm{int}(x) &= \frac{1}{2} \, \bar{\chi}(x) \, g_s \spac \nm A_s(x_-) \, i\np\partial \spac \chi(x)
    \nonumber\\
    &+ \frac{1}{2} \, \bar{\chi}(x) \int_0^1\!ds \, (x-x_-)^\mu \nm^\nu \, g_s \spac F_{s,\mu\nu}(y(s)) \, i\np\partial \spac \chi(x)
    \nonumber\\
    &+ \frac{1}{2} \, \bar{\chi}(x) \int_0^1\!ds \, s \, (x-x_-)^\mu \np^\nu \, g_s \spac F_{s,\mu\nu}(y(s)) \, i\nm\partial \spac \chi(x)
    \nonumber\\
    &+ \bar{\chi}(x) \int_0^1\!ds \, s \, (x-x_-)^\mu \spac g_s \spac F_{s,\mu\nu}(y(s)) \, i\partial^\nu_{\perp} \spac \chi(x) + \text{h.c.} \,,
\end{align}
where $y(s)=x_-+s(x-x_-)$ with $x_\pm^\mu\equiv n_\mp x \, \frac{n_\pm^\mu}{2}$.
To arrive at this form from the one given in~\cite{Beneke:2021aip}, one needs to drop so-called $R$ Wilson lines, which is justified when considering single soft emission only.
As mentioned above, this Lagrangian is exact to all orders in the soft expansion~\cite{Beneke:2002ni}.

As an illustration, we consider the contribution from the second line of~\eqref{eq:lag} to~\eqref{eq:Ai1}. Contracting the soft field-strength tensor and $\bar{\chi}$ with the external state  yields
\begin{align} \label{eq:AII-example}
    -\frac{1}{2} \, g_s \, \bm{T}^a \, k_{[\mu}\varepsilon^\ast_{\nu]}(k) \int_0^1\!ds\, \int\!\frac{d^4q}{(2\pi)^4} \, \frac{i \np q}{q^2} \,  \mathcal{A}_i(q) \,\int\!d^4x\, (x-x_-)^\mu \nm^\nu \, e^{i(p\cdot x+ k\cdot y(s)-q\cdot x)} \,,
\end{align} 
where we evaluated the time-ordered product to a scalar propagator and recognized the non-radiative amplitude in momentum space~\eqref{eq:nr}.
As $x-x_-=x_++x_\perp$, the position-space integral results in 
\begin{align} \label{eq:deltap}
    \int\!d^4x \, (x_++x_\perp)^\mu \, e^{-iq\cdot x} &= i \bigg( \np^\mu \, \frac{\partial}{\partial\np q} + \frac{\partial}{\partial q_{\perp\mu}} \bigg) \, (2\pi)^4 \spac \delta(q) \,.
\end{align}
Integrating by parts, we find
\begin{align} \label{eq:exIBP}
    -\frac{1}{2} \, g_s \, \bm{T}^a \, k_{[\mu}\varepsilon^\ast_{\nu]}(k) \int_0^1\!ds \, \nm^\nu \bigg( \np^\mu \, \frac{\partial}{\partial\np q} + \frac{\partial}{\partial q_{\perp\mu}} \bigg) \bigg[ \frac{\np q}{q^2} \, \mathcal{A}_i(q) \bigg]_{q=\bar{q}(s)} \,,
\end{align} 
where the derivatives are evaluated before $q$ is set to
\begin{align} 
    &\bar{q}^{\spac\mu}(s) = \np(p + s\spac k) \, \frac{\nm^\mu}{2} + \nm k \, \frac{\np^\mu}{2} + s \spac k_\perp^\mu \,.
\end{align}
The denominator becomes $\bar{q}(s)^2= \nm k \, \np p \spac \big(1+s\bar{s} \,\frac{\np k}{\np p}\big)$ with $\bar{s}=1-s$.

Adding the terms obtained from the remaining pieces in~\eqref{eq:lag} and making repeated use of the on-shell conditions $k\cdot\varepsilon=0$ and $k^2=0$ to simplify
\begin{align}
    &k_{[\mu} \varepsilon_{\nu]}^\ast \spac 2 \spac n_\mp^\nu \spac k_\perp^\mu = -k_{[\mu} \varepsilon_{\nu]}^\ast \, n_\mp k \, n_\mp^\nu \spac n_\pm^\mu \,,
    \nonumber\\
    &k_{[\mu} \varepsilon_{\nu]}^\ast \big( \np k \, \nm^\nu + k_\perp^\nu \big) \frac{\partial}{\partial p_{\perp\mu}} = k_{[\mu} \varepsilon_{\nu]}^\ast \, \frac{\nm^\nu \np^\mu}{2} \, k_\perp^\rho \, \frac{\partial}{\partial p_\perp^\rho} \,,
\end{align}
where $a^{[\mu} b^{\nu]} \equiv a^\mu b^\nu-a^\nu b^\mu$, derivatives acting on the non-radiative amplitude can be organized in terms of the orbital angular-momentum operator
\begin{align} \label{eq:angular-momentum}
    L^{\mu\nu} = p^{[\mu}\frac{\partial}{\partial p_{\nu]}} = \np p \, \frac{\nm^{[\mu}}{2} \bigg(\np^{\nu]} \, \frac{\partial}{\partial \np p} + \frac{\partial}{\partial p_{\perp\nu]}} \bigg) \,,
\end{align}
and the momentum-space translation operator 
\begin{align} \label{eq:kdotp}
    k\cdot\frac{\partial}{\partial p} \equiv \np k \, \frac{\partial}{\partial \np p} + k_\perp^\rho \, \frac{\partial}{\partial p_\perp^\rho} \,.
\end{align}
We emphasize again that $\mathcal{A}_i$ does \emph{not} depend on $\nm p$ as it is eliminated by the on-shell condition, which justifies the identification of the right-hand side with the translation operator.
To arrive at the result in terms of these two operators, one has to rearrange terms in the numerator, e.g.
\begin{align}
    2 \np p+ s \spac \np k -s^2 \spac \np k = \np p + \np p \, \bigg( 1+s\bar{s} \,\frac{\np k}{\np p} \bigg) ,
\end{align}
where the bracket cancels $\bar{q}(s)^2$ in the denominator, and reduce squared denominators by
\begin{align} \label{eq:s-derivative-identity}
    &\frac{s-\bar{s}}{\left( 1+s\bar{s} \,\frac{\np k}{\np p} \right)^2}= -\frac{d}{ds} \, \frac{s\bar{s}} {1+s\bar{s} \,\frac{\np k}{\np p}} \,.
\end{align}
Integration by parts in $s$ then transfers the derivative to the non-radiative amplitude. The boundary terms vanish and we find
\begin{align} \label{eq:scalar-intermediate-result}
    \mathcal{A}_{i+1}^{\lag}(p,k) &= -g_s \, \bm{T}^a \, \bigg\{ \frac{\nm\varepsilon^\ast(k)}{\nm k} \, \mathcal{A}_i(p) + \frac{k_\mu\varepsilon^\ast_\nu(k)}{\np p \, \nm k} \int_0^1\!ds
    \nonumber\\
    &\hspace{13mm}\times \bigg[ 2\spac L^{\nu\mu} - \frac{s\bar{s}}{1+s\bar{s} \, \frac{\np k}{\np p}} \, \frac{\nm^{[\nu}\np^{\mu]}}{2}
    \bigg( k\cdot\frac{\partial}{\partial p} - \frac{d}{ds}\bigg) \bigg] \mathcal{A}_i(p+sk) \bigg\} \,.
\end{align}
The second term in the square bracket in~\eqref{eq:scalar-intermediate-result} vanishes by the chain rule.
We then expand in $k$ using
\begin{align} \label{eq:perform-s-integral}
    \int_0^1\!ds \, \mathcal{A}_i(p+sk) = \sum_{n=0}^\infty \int_0^1\!ds \, \frac{s^n}{n!} \, \bigg(k\cdot\frac{\partial}{\partial p}\bigg)^{\!n} \mathcal{A}_i(p) = \sum_{n=0}^\infty \frac{1}{(n+1)!} \, \bigg(k\cdot\frac{\partial}{\partial p}\bigg)^{\!n} \mathcal{A}_i(p) \,,
\end{align}
where the derivative is taken before setting $p_\perp=0$.
This yields the final form for the universal part of the radiative amplitude. 
Summing the $N$ collinear sectors:
\begin{align} \label{eq:finalsQCDuni}
    \mathcal{A}_{N+1}^{\lag}(\{\underline{p}\},k) = -g_s \sum_{i=1}^N \bm{T}_i^a \, \bigg\{ \frac{p_i\cdot\varepsilon^\ast(k)}{p_i\cdot k} + \frac{k_\mu\varepsilon^\ast_\nu(k)}{p_i\cdot k} \sum_{n=0}^\infty \frac{L_i^{\nu\mu}}{(n+1)!} \, \bigg(k\cdot\frac{\partial}{\partial p_i}\bigg)^{\!\!n} \, \bigg\} \,  \mathcal{A}_N(\{\underline{p}\}) \,.
\end{align}
At the $(n+1)$th subleading order in the soft expansion, the universal part of the radiative amplitude is obtained from the non-radiative one by acting with the angular momentum operator and $n$ momentum-space translation operators.
From~\eqref{eq:finalsQCDuni}, it is evident that the universal part of the radiative amplitude is sourced by a single operator.
In SCET, this is reflected by the fact that all Wilson coefficients $C^{Am_1\dots Am_N}$ of power suppressed operators can be obtained from $C^{A0}$. The generalization of~\eqref{eq:finalsQCDuni} to spin-$\frac{1}{2}$ fermions is derived in Appendix~\ref{app:fermions}.

As an aside, we note that while~\eqref{eq:finalsQCDuni} appears to be reparametrization-invariant at each order in the soft expansion, it is actually not due to the definition of $k\cdot\frac{\partial}{\partial p}$ in~\eqref{eq:kdotp}. 
This is because $\mathcal{A}_{N+1}^{\lag}$ is written in terms of the on-shell non-radiative amplitude and solving the on-shell constraint for one of the momentum components inevitably introduces a choice of frame. Reparametrization invariance at each order in the soft expansion is restored when the universal Lagrangian terms are combined with the local terms obtained from on-shell matching of the SCET expression to the full amplitude. 

All-order expressions in the soft expansion have been previously written down in~\cite{Hamada:2018vrw,Li:2018gnc}. In particular, Eq.~(20) of~\cite{Li:2018gnc} agrees with~\eqref{eq:finalsQCDuni} when the gauge symmetry is abelian. In that reference, the ``infinite soft theorem'' was obtained by extracting certain terms from the internal emission amplitude $\mathcal{A}_{N+1}^\mathrm{int}$ by means of the Ward identity. 
The main conceptual difference to the present treatment is that this procedure leaves the remaining pieces unspecified. In contrast, the effective-theory approach presented here provides a direct operator identification of both the universal \emph{and} the process-dependent components, and in consequence enables a systematic classification of the non-universal contributions to which we turn next. 


\section{Degree of non-universality of the local amplitude}
\label{sec:S4_non_uni}

Soft emission directly from the hard vertex described by the amplitude~\eqref{eq:local_emission} is process-dependent in the sense that it cannot be related to the non-radiative amplitude $\mathcal{A}_{N}$, since the associated short-distance coefficients encode information about the hard dynamics of the full radiative process and are determined via matching to the radiative amplitude. 
The symmetries of SCET restrict the structure of the hard-scattering operators sourcing these emissions. Assuming a gauge-invariant and Lorentz-scalar source of the hard process, in this section we decompose the single operator structure identified in~\eqref{eq:Jlocal},~\eqref{eq:js} 
in terms of independent colour-singlet, Lorentz-scalar short-distance coefficients. We compare their number to the corresponding number for the non-radiative process to provide a quantitative measure of the non-universality of soft emission beyond the next-to-soft order.

\subsection{Constraints from gauge invariance}

In this section, we determine the number of colour-singlet structures that source the hard-vertex soft emission for a  configuration of $n_q$ quarks and $\bar{n}_q$ antiquarks ($N=n_q+\bar{n}_q$).
The $SU(3)$ tensor product representation under which the hard-scattering operators~\eqref{eq:Jlocal} transform is 
\begin{align}
    \mathbf{3}^{\otimes n_q}\otimes \bar{\mathbf{3}}^{\otimes \bar{n}_q} \otimes \mathbf{8} \,,
\end{align}
where the octet corresponds to the soft operators~\eqref{eq:js}.
We therefore need to compute the singlet multiplicity, $\mathrm{mult}_{\mathbf{1}}[\mathbf{3}^{\otimes n_q}\otimes \bar{\mathbf{3}}^{\otimes \bar{n}_q}\otimes \mathbf{8}]$, of this representation.

The derivation relies on several group-theory observations.
First, using the decomposition $\mathbf{3}\otimes \bar{\mathbf{3}}=\mathbf{8}\oplus \mathbf{1}$, the original problem can be rewritten in terms of singlet multiplicities of tensor products containing only fundamental and anti-fundamental representations,
\begin{align} \label{eq:8singlets}
    \mathrm{mult}_{\mathbf{1}}\big[ \mathbf{3}^{\otimes n_q}\otimes \bar{\mathbf{3}}^{\otimes \bar{n}_q}\otimes \mathbf{8} \big] = \mathrm{mult}_{\mathbf{1}}\big[ \mathbf{3}^{\otimes n_q+1}\otimes \bar{\mathbf{3}}^{\otimes \bar{n}_q+1}\big] - \mathrm{mult}_{\mathbf{1}}\big[ \mathbf{3}^{\otimes n_q}\otimes \bar{\mathbf{3}}^{\otimes \bar{n}_q}\big] \,.
\end{align}
Second, an irreducible representation $(p,q)\in \mathbf{3}^{\otimes n_q}$ can contribute to a singlet in $\mathbf{3}^{\otimes n_q}\otimes \bar{\mathbf{3}}^{\otimes \bar{n}_q}$ only when combined with its conjugate representation $(q,p)\in \bar{\mathbf{3}}^{\otimes \bar{n}_q} $.
Therefore, the singlet multiplicity can be expressed in terms of the multiplicities of irreducible representations appearing in the tensor products of (anti-)fundamental representations:
\begin{align} \label{eq:3nqbar3barnqsinglets}
    \mathrm{mult}_{\mathbf{1}}\big[\mathbf{3}^{\otimes n_q}\otimes \bar{\mathbf{3}}^{\otimes \bar{n}_q}\big] = \sum_{p,q} \mathrm{mult}_{(p,q)}\big[\mathbf{3}^{\otimes n_q}\big] \, \mathrm{mult}_{(q,p)}\big[\bar{\mathbf{3}}^{\otimes \bar{n}_q}\big] \,,
\end{align}
where the sum is over $p,q\leq \text{min}(n_q,\bar{n}_q)$.
The remaining non-trivial step is the determination of $\mathrm{mult}_{(p,q)}[\mathbf{3}^{\otimes n_q}]$. We obtain this multiplicity using Young diagrams, Young tableaux, and the hook length formula. Details are provided in Appendix~\ref{app:young}.
The final result is
\begin{align} \label{eq:repsinglets}
    \mathrm{mult}_{(p,q)}\big[\mathbf{3}^{\otimes n_q}\big] = (a+b+c)! \,\bigg(\frac{(a+2)! \spac (b+1)! \spac c!}{(a-c+2) (a-b+1) (b-c+1)}\bigg)^{\!-1} \,,
\end{align}
for $a,b,c\,\in \mathbb{N}_0$ fulfilling the constraints
\begin{align} \label{eq:abcconstraints}
    &a= \frac{1}{3}(n_q + 2p + q) \,,
    &b&= \frac{1}{3}(n_q - p + q) \,,
    &c&= \frac{1}{3}(n_q - p - 2q) \,,&a \geq b \geq c\,,
\end{align}
and zero otherwise.
Since $\mathrm{mult}_{(q,p)}\big[\bar{\mathbf{3}}^{\otimes \bar{n}_q}\big] = \mathrm{mult}_{(p,q)}\big[\mathbf{3}^{\otimes \bar{n}_q}\big]$ the same formula can be used for the conjugate representation.

The results are summarized in Table~\ref{tab:color-singlets-soft}, where each entry displays two multiplicities for a configuration  with $n_q$ quarks and $\bar{n}_q$ antiquarks. The entries without parentheses correspond to the number of independent colour-singlet structures in the hard-scattering operator basis $J_\loc^Y$ in the presence of one soft gluon, while the smaller numbers (parenthesized) give the multiplicity for the $(Am)$-type operators~\eqref{eq:collinearcurrents}, which are the ones relevant to the non-radiative amplitude.
As discussed above, before the decomposition, the (anti-) quarks are encoded in collinear gauge-invariant building blocks, while the soft gluon is contained in the soft current~\eqref{eq:js}.
Since all power-suppressed soft currents $J_s$ transform in the adjoint representation of $SU(3)_s$, the number of invariant structures is independent of the order in the soft expansion. It is also independent of the spin of the emitting particle. For fixed $N$, only a small subset of quark–antiquark combinations yields singlets; these satisfy the triality constraint $n_q=\bar{n}_q \; \text{(mod 3)}$.
The corresponding $(n_q,\bar{n}_q)$ pairs are organized along the sub-diagonals of the table.
All configurations with $0\leq n_q,\bar{n}_q\leq5$ are shown. 
The main result of this analysis is that the ratio of invariant amplitudes of the radiative and non-radiative process increases only modestly. For example, for a hard process with two quarks and two antiquarks, the number of colour-singlet amplitudes is two for the non-radiative process and only two more are required to describe the radiative process.

\begin{table}[t]
    \centering
    \renewcommand{\arraystretch}{1.25}
    \setlength{\tabcolsep}{7pt}
    \arrayrulecolor{rulegray}
    {
    $\displaystyle
        \operatorname{mult}_{\mathbf 1}\!
        \big[
            \mathbf 3^{\otimes n_q}
            \otimes
            \bar{\mathbf 3}^{\otimes \bar n_q}
            \otimes
            \mathbf 8
        \big]$
    }\par\vspace{6pt}

    \begin{tabular}{|>{\columncolor{headergray}}c|*{6}{c|}}
        \hline
        \rowcolor{headergray}
        \rule{0pt}{2.6ex}
        \boldmath$n_q\backslash\bar n_q$
            & \bfseries 0
            & \bfseries 1
            & \bfseries 2
            & \bfseries 3
            & \bfseries 4
            & \bfseries 5 \\
        \hline
        \bfseries 0
            & \pair{0}{-}
            & \pair{0}{0}
            & \pair{0}{0}
            & \pair{2}{1}
            & \pair{0}{0}
            & \pair{0}{0} \\
        \hline
        \bfseries 1
            & \pair{0}{0}
            & \pair{1}{1}
            & \pair{0}{0}
            & \pair{0}{0}
            & \pair{8}{3}
            & \pair{0}{0} \\
        \hline
        \bfseries 2
            & \pair{0}{0}
            & \pair{0}{0}
            & \pair{4}{2}
            & \pair{0}{0}
            & \pair{0}{0}
            & \pair{36}{11} \\
        \hline
        \bfseries 3
            & \pair{2}{1}
            & \pair{0}{0}
            & \pair{0}{0}
            & \pair{17}{6}
            & \pair{0}{0}
            & \pair{0}{0} \\
        \hline
        \bfseries 4
            & \pair{0}{0}
            & \pair{8}{3}
            & \pair{0}{0}
            & \pair{0}{0}
            & \pair{80}{23}
            & \pair{0}{0} \\
        \hline
        \bfseries 5
            & \pair{0}{0}
            & \pair{0}{0}
            & \pair{36}{11}
            & \pair{0}{0}
            & \pair{0}{0}
            & \pair{410}{103} \\
        \hline
    \end{tabular}
    \caption{Number of independent colour-singlet operators for a process with $n_q$ quarks and $\bar{n}_q$ antiquarks. The main entries correspond to the hard-scattering operators in~\eqref{eq:Jlocal} for the radiative process containing one additional soft gluon, while the smaller numbers in parentheses correspond to the $(Am)$-type operators in~\eqref{eq:collinearcurrents} without the octet in the tensor product. Entries with fixed $N=n_q+\bar n_q$ lie along the sub-diagonals. In both cases, the multiplicity is independent of the order in the soft expansion, and non-vanishing multiplicities satisfy the triality constraint $n_q=\bar n_q\pmod 3$.}
    \label{tab:color-singlets-soft}
\end{table}

\subsection{Constraints from Lorentz invariance}

We now turn to the corresponding counting of Lorentz scalars assuming a \emph{Lorentz-scalar} source. As we are considering single soft emissions only, the soft operators~\eqref{eq:js} always appear inside a one-gluon matrix element 
\begin{align} \label{eq:newjsmatrix}
    \langle k | J_s^{(n+2),\rho_1\dots\rho_{n}\mu\nu }(0) |0\rangle = -g_s \spac k^{\rho_1}\dots k^{\rho_{n}} k^{[\mu} \varepsilon^{\ast\nu]}(k) \,.
\end{align}
One thus deals with a tensor $A^{\rho_1\dots\rho_n,\mu\nu}$ transforming in $\mathrm{Sym}^n\otimes \Lambda^2$ where $\mathrm{Sym}^k$ ($\Lambda^k$) is the space of all symmetric (anti-symmetric) tensors of rank $k$.
The dimension of the underlying space is $D=\min\{N,4\}$ in four space-time dimensions.
The exception is processes without a source, and hence with both incoming and outgoing hard particles, as momentum conservation makes one of the vectors $\nm[i]$ linearly dependent.
In this case, $D=\min\{N-1,4\}$.
These considerations lead to the naive number
\begin{align} \label{eq:Lorentz_scalars_naive}
    \binom{D+n-1}{n} \binom{D}{2} \leq (n+1) (n+2) (n+3)
\end{align}
of independent Lorentz scalars. However, additional constraints arise from the Bianchi identities and tracelessness, because contracting any two indices in~\eqref{eq:newjsmatrix} yields zero due to the on-shell conditions $k^2=0$ and $k\cdot\varepsilon=0$. Two cases need to be distinguished.

For $D\leq3$, it is not possible to span the whole Minkowski space by a set of independent $\nm[i]$ vectors.
In this case, the tracelessness constraints involve directions orthogonal to the $\nm[i]$ and do not reduce the number of independent Lorentz scalars.
The Bianchi identities can be formulated as 
\begin{align} \label{eq:Bianchi}
    B_{\rho_1,\mu\nu}^{\rho_2\dots\rho_n} \equiv A^{\rho_1\rho_2\dots\rho_n,\mu\nu} - A^{\mu\rho_2\dots\rho_n,\rho_1\nu} - A^{\nu\rho_2\dots\rho_n,\mu\rho_1} = 0 \,.
\end{align}
The tensor $B_{\rho_1,\mu\nu}^{\rho_2\dots\rho_n}$ transforms in $\mathrm{Sym}^{n-1}\otimes\Lambda^3$ and thus the number of independent Lorentz scalars is reduced to
\begin{align} \label{eq:LorentzcountingD=3}
    \binom{D+n-1}{n} \binom{D}{2} - \binom{D+n-2}{n-1} \binom{D}{3} =
    \begin{cases}
        \, n+1 \,, & D=2 \,, \\[1mm]
        \, (n+1)(n+3) \,, & D=3 \,.
    \end{cases}
\end{align}
For $D\leq3$ the large $n$ behaviour is the same as for the naive counting.

Next, we study $D=4$. In this case, any four of the $\nm[i]^\mu$ vectors span the full Minkowski space, and can be used to express the metric and epsilon tensors as
\begin{align}
    g^{\mu\nu} &= \sum_{i,j} (G^{-1})_{ij} \, \nm[i]^\mu \nm[j]^\nu \,,
    &
    \epsilon^{\mu\nu\rho\sigma} &= \sum_{i,j,k,l} \frac{\epsilon_{ijkl}}{\sqrt{-\det G}} \, \nm[i]^\mu \nm[j]^\nu \nm[k]^\rho \nm[l]^\sigma \,,
\end{align}
with Gram matrix $G_{ij}=\nm[i]\nm[j]$.
We ignore exceptional phase-space configurations where these four $\nm[i]^\mu$ are linearly dependent.
This has two important consequences: First, not all Bianchi identities in~\eqref{eq:Bianchi} are independent. The tensor $\epsilon_{\rho_1\rho_2\mu\nu}B_{\rho_1,\mu\nu}^{\rho_2\dots\rho_n}$ vanishes identically. As it transforms in $\mathrm{Sym}^{n-2}$, the number of constraints is reduced by $\binom{4+n-3}{n-2}=\binom{n+1}{3}$. 
Second, tracelessness gives constraints. They can be formulated as 
\begin{align} \label{eq:Trace}
    T_\nu^{\rho_2\dots\rho_n} \equiv g_{\rho_1\mu} \, A^{\rho_1\rho_2\dots\rho_n,\mu\nu} = 0 \,,
\end{align}
which are $4\binom{4+n-2}{n-1}=4\binom{n+2}{3}$ equations. Traces involving two $\rho_i$ are already included by the Bianchi identities
\begin{align}
    g_{\rho_1\rho_2} \, A^{\rho_1\rho_2\dots\rho_n,\mu\nu} = g_{\rho_1\rho_2} \, A^{\mu\rho_2\dots\rho_n,\rho_1\nu} + g_{\rho_1\rho_2} \, A^{\nu\rho_2\dots\rho_n,\mu\rho_1} = T_\nu^{\mu\rho_3\dots\rho_n} - T_\mu^{\nu\rho_3\dots\rho_n} \,.
\end{align}
However, there is a redundancy in~\eqref{eq:Trace} which is
\begin{align}
    g_{\rho_2\nu} \, T_\nu^{\rho_2\dots\rho_n} = g_{\rho_1\mu} \, g_{\rho_2\nu} \, A^{\rho_1\rho_2\dots\rho_n,\mu\nu} = 0 \,.
\end{align}
As the left-hand side transforms in $\mathrm{Sym}^{n-2}$, the number of constraints is again reduced by $\binom{n+1}{3}$. Putting everything together, we find for $D=4$
\begin{align}
    \binom{n+3}{n} \binom{4}{2} - \binom{n+2}{n-1} \binom{4}{3} - 4\,\binom{n+2}{3} + 2\,\binom{n+1}{3} = 2\spac (n+1)(n+3)
\end{align}
independent Lorentz scalars.
We note that the large-$n$ behaviour is reduced from the naive $n^3$ in~\eqref{eq:Lorentz_scalars_naive} to $n^2$. Unlike the case of colour singlets, the number of Lorentz scalars grows with the order $n+2$ of the soft expansion, but does not depend on $N$ if $N>4$.
The total degree of non-universality of soft emissions beyond the next-to-soft order is then given by the product of the number of gauge singlets and Lorentz scalars.

As an illustration, we consider the process $u(p_1)+d(p_2) \to u(p_3)+d(p_4)+g(k)$ in scalar QCD. The universal part of the amplitude is obtained from~\eqref{eq:finalsQCDuni} by matching onto the leading-power non-radiative amplitude.
The latter contains two invariant colour structures, the singlet $(\bar{\chi}_3 \chi_1)(\bar{\chi}_4\chi_2)$ and the octet $(\bar{\chi}_3 t^a \chi_1)(\bar{\chi}_4 t^a \chi_2)$.
Process-dependent contributions are extracted by subtracting the universal terms from the soft expansion of the full-theory amplitude and matching the remainder onto a basis of colour-singlet hard-scattering operators.
For this process, the basis consists of four colour-singlet operators
\begin{align}
    &(\bar{\chi}_{3} \spac \chi_{1})(\bar{\chi}_{4} \spac J_s^Y \chi_{2}) \,,
    &
    &(\bar{\chi}_{3}\chi_{2})(\bar{\chi}_{4} \spac J_s^Y \chi_{1}) \,,
    &
    &(\bar{\chi}_{4} \spac \chi_{1})(\bar{\chi}_{3} \spac J_s^Y \chi_{2}) \,,
    &
    &(\bar{\chi}_{4}\chi_{2})(\bar{\chi}_{3} \spac J_s^Y \chi_{1}) \,,
\end{align}
of which only two contribute at tree-level due to the $t$-channel topology.
This matching procedure determines the complete set of Wilson coefficients at each order in the soft expansion.
According to~\eqref{eq:LorentzcountingD=3} there are three independent Lorentz scalars for $J_s^Y$ at next-to-next-to-soft order ($n=0$) for an $N=4$ process without source $D=3$.
They can be chosen as
\begin{align}
    &\nm[1]^\mu \nm[2]^\nu \spac F_{s,\mu\nu} \,,
    &
    &\nm[1]^\mu \nm[3]^\nu \spac F_{s,\mu\nu} \,,
    &
    &\nm[2]^\mu \nm[3]^\nu \spac F_{s,\mu\nu} \,,
\end{align}
where $\nm[4]^\mu$ is eliminated by momentum conservation. Thus, the total number of non-universal amplitudes required to describe the radiative process is twelve compared to the two non-radiative amplitudes.


\section{Summary}
\label{sec:S5_Summary}

In this article, we investigated the single soft-gluon emission amplitude to all orders in the soft expansion in the framework of on-shell soft-collinear effective theory. In this framework, the amplitude splits automatically into a Lagrangian emission amplitude, which is universal, and emission from the hard process, where both are separately gauge-invariant~\cite{Beneke:2021umj}. The Lagrangian emission part results 
in the compact expression~\eqref{eq:finalsQCDuni}, which is the natural all-order extension of the LBK amplitude, and has previously been obtained in a different approach~\cite{Li:2018gnc}. The local-emission amplitude is written in terms of a single gauge-invariant operator structure to all orders in the soft expansion. Its matching coefficient defines the non-universal part of the amplitude in the sense that it cannot be expressed in terms of the non-radiative amplitude. Still, the matching coefficients are well-defined universal objects on their own, and decomposing them into colour singlets and Lorentz scalars, one can enumerate the ``non-universal'' amplitudes given the colour and spin of the particles of the non-radiative amplitude. The corresponding results were obtained explicitly for scalars in the (anti-)triplet $SU(3)$ representation.  

\subsubsection*{Note added}

In the final stage of writing this article, Eq.~\eqref{eq:finalsQCDuni} was also obtained (in a different way) from SCET in~\cite{Cohen:2026xnr}.

\paragraph{Acknowledgments}

This research was supported by the Excellence Cluster ORIGINS, which is funded by the Deutsche Forschungsgemeinschaft (DFG, German Research Foundation) under Germany’s Excellence Strategy -- EXC 2094 -- 390783311.


\begin{appendix}


\section{Fermions}
\label{app:fermions}

In this section, we summarize the results of the preceding analysis for the emission of true spin-$\frac{1}{2}$ quarks and antiquarks rather than scalar ones.  For particles with spin, the total angular-momentum operator appearing in the soft theorem~\eqref{eq:LBK} contains both the orbital contribution $L^{\mu\nu}$ and the spin contribution $\Sigma^{\mu\nu}$. In SCET, the small component of the collinear quark field is integrated out, while the remaining field satisfies the projection condition $\nms[i]\chi_i=0$. When acting on a collinear spinor, the spin operator can therefore be decomposed as
\begin{align} \label{eq:spin-operator}
    \bar{\xi}(p) \, \Sigma^{\mu\nu} 
    = \bar{\xi}(p) \, \frac{1}{4}\big[\gamma^\mu,\gamma^\nu\big] 
    = \bar{\xi}(p) \, \frac{1}{4} \bigg( [\gamma_\perp^\mu,\gamma_\perp^\nu] -\nps\gamma_\perp^{[\mu}\nm^{\nu]} - \np^{[\mu}\nm^{\nu]} \bigg) \,.
\end{align}
We denote the three contributions by $\Sigma_\perp^{\mu\nu}$, $\Sigma_{\perp+}^{\mu\nu}$ and $\Sigma_{+-}^{\mu\nu}$, respectively.

The presence of the external spinors modifies the left-hand sides of~\eqref{eq:nr} and~\eqref{eq:tproduct_uni}, which must be multiplied by $\prod_{i=1}^N\bar{\xi}_i(p_i)$. Since part of the collinear field is integrated out, the full-theory spinor is non-trivially related to the  SCET collinear spinor. 
This, in turn, yields the following matching equation between the amplitude $\hat{\mathcal{A}}$ stripped of collinear spinors and the amplitude $\mathcal{A}$ stripped of full-theory spinors:
\begin{align} \label{eq:spinor-relation}
    \bar{u}(p_i) &= \bar{\xi}_i(p_i) \bigg( 1-\frac{\slashed{p}\vphantom{p}_{i\perp}}{\np[i]p_i} \frac{\nps[i]}{2} \bigg)
    &&\Rightarrow&
    \hat{\mathcal{A}}(\{\underline{p}\}) &= \prod_{i=1}^N \bigg( 1-\frac{\slashed{p}\vphantom{p}_{i\perp}}{\np[i]p_i} \, \frac{\nps[i]}{2} \bigg) \mathcal{A}(\{\underline{p}\}) \,.
\end{align}
The Dirac indices of the brackets in the right expression contract separately with the respective external leg of $\mathcal{A}$. In principle, one has to allow for additional terms proportional to any of the $\nms[i]$ on the right-hand side.
However, these terms will not contribute.

As in the scalar case, the amplitude is split into the universal Lagrangian emission and local emission part. We focus on the former. Omitting the index $i$, the interaction Lagrangian is given by~\cite{Beneke:2002ni}
\begin{align}
    \mathcal{L}^\mathrm{int}(x) &= \bar{\chi}(x) \, g_s \spac \nm A_s(x_-) \, \frac{\nps}{2} \, \chi(x)
    \nonumber\\
    &+ \bar{\chi}(x) \int_0^1\!ds \, (x-x_-)^\mu \nm^\nu \, g_s \spac F^s_{{\mu}{\nu}}(y(s)) \, \frac{\nps}{2} \, \chi(x)
    \nonumber\\
    &+\bar{\chi}(x) \int_0^1\!ds \, s \, (x-x_-)^\mu\gamma_\perp^{\nu} \, g_s \spac F^{s}_{{\mu}{\nu}}(y(s)) \, \frac{i\slashed{\partial}_\perp}{i\np \partial} \, \frac{\nps}{2} \, \chi(x) + \text{h.c.}
    \nonumber\\[-1mm]
    &-\bigg(\frac{i\slashed{\partial}_\perp}{i\np\partial} \,\bar{\chi}(x)\bigg) \int_0^1\!ds \, s \, (x-x_-)^\mu \np^\nu \, g_s \spac F^s_{{\mu}{\nu}}(y(s)) \, \frac{i \slashed{\partial}_\perp}{i\np \partial} \, \frac{\nps}{2} \, \chi(x) \,.
\end{align}
The computation of the universal terms~\eqref{eq:Ai1} proceeds analogously to the scalar case.
In this case, one does not only identify the angular momentum operator~\eqref{eq:angular-momentum}, but also certain components of the spin operator~\eqref{eq:spin-operator}.
The radiative amplitude then takes the form 
\begin{align}
    \mathcal{A}_{N+1}^{\lag}(\{\underline{p}\},k) &= -g_s \sum_{i=1}^N \bm{T}_i^a \, \bigg\{ \frac{\nm[i]\varepsilon^\ast(k)}{\nm[i] k} \, \hat{\mathcal{A}}_N(p) + \frac{2 \spac k_\mu\varepsilon^\ast_\nu(k)}{\np[i] p_i \, \nm[i] k} \int_0^1\!ds
    \nonumber\\*
    &\hspace{13mm}\times \Big[ L_i^{\nu\mu} + (\Sigma_{i\perp}^{\nu\mu} + \Sigma_{i+-}^{\nu\mu}) \, \delta(\bar{s}) \Big] \hat{\mathcal{A}}_N(p_i+sk) \bigg\} \,,
\end{align}
where the $\delta(\bar{s})$ is a boundary term from integration by parts.
Expanding in $k$ similarly to~\eqref{eq:perform-s-integral}, we find the final result
\begin{align} \label{eq:uni_final_fermions}
    \mathcal{A}_{N+1}^{\lag}(\{\underline{p}\},k) =& -g_s \sum_{i=1}^N \bm{T}_i^a \, \bigg\{ \frac{p_i\cdot\varepsilon^\ast(k)}{p_i\cdot k} + \frac{k_\mu\varepsilon^\ast_\nu(k)}{p_i\cdot k}
    \nonumber\\*
    &\times \sum_{n=0}^\infty \bigg[ \frac{L_i^{\nu\mu}}{(n+1)!} + \frac{1}{n!} \, (\Sigma_{i\perp}^{\nu\mu} + \Sigma_{i+-}^{\nu\mu}) \bigg] \bigg(k\cdot\frac{\partial}{\partial p_i}\bigg)^{\!n} \,\bigg\} \, \hat{\mathcal{A}}_N(\{\underline{p}\}) \,.
\end{align}
As expected, for fermions the universal soft operator consists of the same orbital contribution as in the scalar case, supplemented by spin-dependent terms.
Note that the mixed transverse-longitudinal component of the spin operator $\Sigma_{i\perp+}^{\nu\mu}$ does not appear explicitly in the SCET universal radiative amplitude. 
At the next-to-soft order ($n=0)$, one recovers it and the well known result~\eqref{eq:LBK} by using relation~\eqref{eq:spinor-relation} and observing that
\begin{align}
    \Big[ L_i^{\nu\mu}, \bigg(1-\frac{\slashed{p}\vphantom{p}_{i\perp}}{\np[i]p_i} \, \frac{\nps[i]}{2} \bigg) \Big]_{p_{i\perp}=0} = \Sigma_{i\perp +}^{\nu\mu} \,.
\end{align}
The additional terms proportional to $\nms[i]$ in~\eqref{eq:spinor-relation} vanish because $\bar{\xi}_i\spac\nms[i]=0$.
The next-to-soft expression was already derived from SCET in~\cite{Beneke:2021umj}.
For the same reason as discussed in the main text, our result~\eqref{eq:uni_final_fermions} is not reparametrization-invariant at a given order in the soft expansion, but here the non-invariance is seen more directly in the spin term. 


\section{\texorpdfstring{\boldmath $C^{Am_1\dots Am_N}$}{Cᴬᴹ} coefficients}
\label{app:CAM}

Our starting point to derive the relation between $C^{Am_1\dots Am_N}$ and $C^{A0}$ is~\eqref{eq:nr}. Lorentz invariance requires that $\mathcal{A}_N$ is a function of the $\binom{N}{2}$ invariants $s_{ij}=2p_i\cdot p_j$ only.
At the same time, boost invariance requires that the $C^{Am_1\dots Am_N}$ are functions of
\begin{align}
    s_{ij}^{--} = 2 \spac p_{i-} p_{j-} = \frac{1}{2} \, \nm[i]\nm[j] \spac \np[i] p_i \spac \np[j] p_j
\end{align}
only. To reconcile these two statements, all $C^{Am_1\dots Am_N}$ coefficients need to be related to $C^{A0}$, which in turn relates to the non-radiative amplitude by
\begin{align} \label{eq:CA0_Anr}
    C^{A0}(\{\underline{s^{--}}\}) = \mathcal{A}_N(\{\underline{s}\}) \big|_{p_{i\perp}=0} = \mathcal{A}_N(\{\underline{s^{--}}\}) \,.
\end{align}
For the following discussion, it is convenient to introduce
\begin{align}
    s_{ij}^{-\perp} &= 2 \spac p_{i-}p_{j\perp} + 2p_{j-}p_{i\perp} = \np[i] p_i \spac \nm[i] p_{j\perp} + \np[j] p_j \spac \nm[j] p_{i\perp} \,,
    \nonumber\\
    s_{ij}^{\perp\perp} &= 2 \spac p_{i\perp}p_{j\perp} \,,
    \nonumber\\[-1mm]
    s_{ij}^{-+} &= 2p_{i-}p_{j+} + 2p_{j-}p_{i+} = -\frac{1}{2} \bigg( \frac{\np[i] p_i}{\np[j] p_j} \, p_{j\perp}^2 \spac \np[j]\nm[i] + \frac{\np[j] p_j}{\np[i] p_i} \, p_{i\perp}^2 \spac \np[i]\nm[j] \bigg)  \,,
    \nonumber\\[-2mm]
    s_{ij}^{+\perp} &= 2\spac p_{i+}p_{j\perp} + 2p_{j+}p_{i\perp} = - \bigg( \frac{p_{i\perp}^2}{\np[i] p_i} \, \np[i] p_{j\perp} + \frac{p_{j\perp}^2}{\np[j] p_j} \, \np[j] p_{i\perp} \bigg) \,,
    \nonumber\\[-2mm]
    s_{ij}^{++} &= 2 \spac p_{i+}p_{j+} = \frac{1}{2} \, \frac{p_{i\perp}^2}{\np[i] p_i} \, \frac{p_{j\perp}^2}{\np[j] p_j} \, \np[i]\np[j] \,.
\end{align}
These are the light-cone decomposed versions of $s_{ij}$ which scale as $(\lambda,\lambda^2,\lambda^2,\lambda^3,\lambda^4)$, respectively.
Expanding the left-hand side of~\eqref{eq:nr} and comparing order by order in $\lambda$ to the right-hand side yields
\begin{align} \label{eq:master_CAM}
    & \sum_{|M|} \, (p_{1\perp})^{m_1} \dots (p_{N\perp})^{m_N} \, C^{Am_1\dots Am_N} = \quad \sum_{\mathclap{\substack{b_1,b_2,b_3,b_4 \\[1mm] b_1+2b_2+3b_3+4b_4= |M|}}} \quad \frac{1}{b_1!\spac b_2! \spac b_3! \spac b_4!} \bigg(\frac{1}{2} \sum_{(ij)} s_{ij}^{-\perp} \, \frac{\partial}{\partial s_{ij}^{--}} \bigg)^{\!b_1}
    \\*\nonumber
    &\qquad \times \bigg(\frac{1}{2} \sum_{(ij)} (s_{ij}^{\perp\perp} + s_{ij}^{-+}) \, \frac{\partial}{\partial s_{ij}^{--}} \bigg)^{\!b_2} \bigg(\frac{1}{2} \sum_{(ij)} s_{ij}^{+\perp} \, \frac{\partial}{\partial s_{ij}^{--}} \bigg)^{\!b_3} \bigg(\frac{1}{2} \sum_{(ij)} s_{ij}^{++} \, \frac{\partial}{\partial s_{ij}^{--}} \bigg)^{\!b_4} \, C^{A0} \,,
\end{align}
where the sums $(ij)$ are over all unordered pairs with $i\neq j$. From this result, explicit relations for given $|M|=\sum_i m_i$ are easily derived.

For $|M|=1$, the constraint on the right-hand side is only solved by $\vec{b}=(1,0,0,0)$. Therefore, the formula simplifies drastically
\begin{align}
    \sum_{i=1}^N p_{i\perp}^\mu C_{i,\mu}^{A1} = \frac{1}{2} \sum_{(ij)} s_{ij}^{-\perp} \, \frac{\partial C^{A0}}{\partial s_{ij}^{--}} = \sum_{(ij)} \frac{2}{\np[i] p_i} \, \nm[j] p_{i\perp} \, \frac{\partial C^{A0}}{\partial\nm[i]\nm[j]} \,,
\end{align}
where we applied the chain rule and exploited the symmetry under $i\leftrightarrow j$.
Comparing the coefficient of $p_{i\perp}$ on both sides yields
\begin{align}
    C_i^{A1,\mu} = \frac{2}{\np[i] p_i} \sum_{j\neq i} \nm[j]^{\mu_{\perp i}} \, \frac{\partial C^{A0}}{\partial \nm[i]\nm[j]} \,.
\end{align}
This result matches the known one~\cite{Beneke:2021umj} when accounting for the difference $\pm i\partial_{i\perp} \chi_i$ in the definition of subleading-power building blocks and using the chain rule (in a suggestive but mathematically obscure way).
In the special case $N=2$, one usually works in the back-to-back frame where $n_\pm^\mu\equiv n_{1\pm}^\mu=n_{2\mp}^\mu$ and then $C^{A0,A1}=C^{A1,A0}=0$.

As a second example, we look at $|M|=2$. In this case, there are two solutions to the constraint on the sum in~\eqref{eq:master_CAM}, $\vec{b}=(2,0,0,0)$ and $\vec{b}=(0,1,0,0)$.
Summing both contributions and comparing coefficients, we find
\begin{align}
    C_i^{A2,\mu\nu} = \frac{2}{(\np[i] p_i)^2} \sum_{j,k\neq i} \nm[j]^{\mu_{\perp i}} \nm[k]^{\nu_{\perp i}} \, \frac{\partial^2 C^{A0}}{\partial \nm[i]\nm[j]\partial \nm[i]\nm[k]} - \frac{g_{\perp i}^{\mu\nu}}{(\np[i] p_i)^2} \sum_{j\neq i} \np[i]\nm[j] \, \frac{\partial C^{A0}}{\partial \nm[i]\nm[j]}
\end{align}
in agreement with \cite{Cohen:2026xnr} and 
\begin{align} \label{eq:A1A1}
    C_{ij}^{A1A1,\mu\nu} = \frac{4}{\np[i] p_i \spac \np[j] p_j} \sum_{k\neq i} \sum_{l\neq j} \nm[k]^{\mu_{\perp i}} \nm[l]^{\nu_{\perp j}} \, \frac{\partial^2 C^{A0}}{\partial \nm[i]\nm[k]\partial \nm[j]\nm[l]} + \frac{4 \spac g^{\mu_{\perp i}\nu_{\perp j}}}{\np[i] p_i \spac \np[j] p_j} \, \frac{\partial C^{A0}}{\partial \nm[i]\nm[j]} \,,
\end{align}
where $C_{ij}^{A1A1}$ is a shorthand notation for $C^{Am_1\dots Am_N}$ with only $m_i=m_j=1$ being non-zero.
For $N=2$ and in the back-to-back frame, $C^{A0,A2}$ and $C^{A2,A0}$ vanish but one has after applying the chain rule
\begin{align}
    C^{A1A1,\mu\nu} = 2 \spac g_\perp^{\mu\nu} \, \frac{\partial C^{A0}}{\partial Q^2} \,,
\end{align}
with $Q^2=\np p_1 \spac \nm p_2$. This was recently confirmed at the one-loop order~\cite{Beneke:2026ogs}.

If one is interested in the coefficients $C_i^{Am_i}$ with only a single misaligned direction, formula~\eqref{eq:master_CAM} can be simplified to
\begin{align} \label{eq:CAmi}
    (p_{i\perp})^{m_i} \, C_i^{Am_i} = \;\sum_{\mathclap{\substack{b_1,b_2 \\[0.5mm] b_1+2b_2= m_i}}} \; \frac{1}{b_1!\spac b_2!} \bigg(\frac{1}{2} \sum_{(jk)} s_{jk}^{-\perp} \, \frac{\partial}{\partial s_{jk}^{--}} \bigg)^{\!b_1} \bigg(\frac{1}{2} \sum_{(jk)} s_{jk}^{-+} \, \frac{\partial}{\partial s_{jk}^{--}} \bigg)^{\!b_2} \, C^{A0} \Big|_{p_{j\perp}=0} \,,
\end{align}
where we set all $p_{j\perp}=0$ for $j\neq i$ on the right-hand side.

For spin-$\frac{1}{2}$ fermions, the $C_i^{Am_i}$ coefficients acquire an additional spin term and take the form
\begin{align}
    C_i^{Am_i}=C^{Am_i}_{i,\mathrm{orb}}-\frac{\gamma_{\perp i}}{\np[i]p_i} \, \frac{\nps[i]}{2} \, C^{A(m_i-1)}_{i,\mathrm{orb}} \,,
\end{align}
where $C^{Am_i}_{i,\mathrm{orb}}$ corresponds to the coefficient presented in~\eqref{eq:CAmi}. In the second term on the right-hand side, it is understood that the (unwritten) Lorentz index of $\gamma_{\perp i}$ is symmetrized (including $1/m_i!$) with the $m_i-1$ Lorentz indices of $C^{A(m_i-1)}_{i,\mathrm{orb}}$.


\section{Multiplicities of irreps from \texorpdfstring{\boldmath $SU(3)$}{SU(3)} fundamentals}
\label{app:young}

In this appendix, we determine the multiplicity $\mathrm{mult}_{(p,q)}[\mathbf{3}^{\otimes n}]$ with which the $SU(3)$ irreducible representation $(p,q)$, with $p,q\in\mathbb{N}_0$, appears in $\mathbf{3}^{\otimes n}$:
\begin{align}
    \mathbf{3}^{\otimes n}
    =
    \underbrace{(1,0)\otimes\dots\otimes(1,0)}_{n\ \text{times}}
    =
    \bigoplus_{p,q}
    \mathrm{mult}_{(p,q)}\big[\mathbf{3}^{\otimes n}\big]\,(p,q) \,.
\end{align}
To determine these multiplicities, we first recall that the decomposition of the tensor product of an arbitrary $SU(3)$ representation $(p,q)$ with the fundamental representation $(1,0)$ is given by
\begin{align} \label{decompo}
    (p,q)\otimes(1,0)
    =
    \underbrace{(p+1,q)}_{\text{symmetrization}}
    \oplus
    \underbrace{(p-1,q+1)}_{\epsilon\text{-contraction}}
    \oplus
    \underbrace{(p,q-1)}_{\text{trace}} \,,
\end{align}
where the $\epsilon$-contraction (the trace) is absent for $p=0$ ($q=0$).
The representation $(p,q)$ can be realized as a tensor
$T^{\mu_1 \ldots \mu_p}_{\nu_1 \ldots \nu_q}$ with $p$ upper and $q$ lower indices.
The three terms in~\eqref{decompo} correspond, respectively, to symmetrizing the new index with the upper indices, contracting with the $\epsilon$-tensor, and performing a trace.

Assume that, in order to obtain a given representation $(p,q)$ from $\mathbf{3}^{\otimes n}$, we perform $a$ symmetrizations, $b$ $\epsilon$-contractions, and $c$ traces.
Then one may write
\begin{align}
    \mathbf{3}^{\otimes n} \supset (a-b,b-c) \equiv (p,q) \,.
\end{align}
This yields the constraints given in~\eqref{eq:abcconstraints} on $a,b,c$.
Each irreducible representation $(p,q)$ is therefore associated with a unique triple $(a,b,c)$.
The multiplicity of $(p,q)$ in $\mathbf{3}^{\otimes n}$ arises from the number of distinct orderings in which the corresponding symmetrizations, $\epsilon$-contractions, and traces can be performed. To encode this, we introduce three ordered sequences
\begin{align}
    (A_1,\dots,A_a), \qquad (B_1,\dots,B_b), \qquad (C_1,\dots,C_c) \,,
\end{align}
where each entry specifies the step at which a given operation is performed. Since every one of the $n$ steps is exactly one of the three operations, the
sequences together form a partition of $\{1,\dots,n\}$. The entries in each sequence are ordered increasingly.
In particular, $A_1=1$, since the first step is always $(0,0)\otimes(1,0)=(1,0)$.
We arrange these sequences into a Young diagram of shape $(a,b,c)$:
\begin{align}
  \scalebox{1}{$\ytableausetup{boxsize=7mm}
    \begin{ytableau}
        A_1 & \cdots & A_c & \cdots & A_b & \cdots & A_a
        \\
        B_1 & \cdots & B_c & \cdots & B_b
        \\
        C_1 & \cdots & C_c
    \end{ytableau}
  $}
\end{align}
Since $p,q\geq0$ at every intermediate step, each $\epsilon$-contraction must be preceded by a further symmetrization and each trace by a further $\epsilon$-contraction, so the columns are strictly ordered, $A_i<B_i<C_i$, and valid fillings correspond precisely to standard Young tableaux.
It follows that the multiplicity of $(p,q)$ in $\mathbf{3}^{\otimes n}$ is equal to the number of standard Young tableaux of shape $(a,b,c)$.
This number, which is exactly the one appearing in the final result~\eqref{eq:repsinglets}, is given by the hook-length formula for the Young diagram of shape $(a, b,c)$ associated with the partition $(a\geq b\geq c)$ of $n=a+b+c$.

\end{appendix}

\pdfbookmark[1]{References}{Refs}
\bibliography{references}

\end{document}